\documentclass[11pt]{article}
\usepackage{graphicx}
\usepackage{subfigure}
\usepackage{caption}
\usepackage{float}
\usepackage{amsmath}
\usepackage{amssymb}
\usepackage{amsxtra}

\usepackage{hyperref}

\title{Neutrino cooling effect of primordial hot areas in dependence on its size}
\author{{K. M. Belotsky} $^{1}$, M. M. El Kasmi $^{1,2}$, S. G. Rubin $^{1,3}$\\ $^{1}$ National Research Nuclear University MEPhI,\\ Kashirskoe Shosse 31, Moscow 115409, Russia\\ 
e-mail k-belotsky@yandex.ru\\ $^{2}$Physics Department, Faculty of Science, Sohag University,\\ Sohag Center, Sohag 82524, Egypt\\ e-mail m.elkasemy@science.sohag.edu.eg\\ $^{3}$ N.~I.~Lobachevsky Institute of Mathematics and Mechanics,\\ Kazan  Federal  University, \mbox{Kremlevskaya  Street 18},\\ Kazan 420008, Russia\\ e-mail sgrubin@mephi.ru}

\date{September 2020}

\begin{document}
\maketitle

\begin{abstract}
We consider the temperature dynamics of hypothetical primordial hot areas in the Universe.  Such areas can be produced by the primordial density inhomogeneities and can survive to the modern era, in particular due to primordial black hole (PBH) cluster of size $R \gtrsim 1$ pc and more. Here we concentrate on the neutrino cooling effect which is realized due to reactions of weak $p\leftrightarrow n$ transitions and $e^{\pm}$ annihilation. 
%
The given neutrino cooling mechanism is found to work in a wide range of parameters. For those parameters typical for PBH cluster considered, the cooling mechanism is quite valuable for the temperatures $T \gtrsim 3$ MeV.
\end{abstract}

\noindent Primordial hot areas, primordial black holes, cosmic neutrinos

\section{Introduction}

There are some observations \cite{alihaimoud2019electromagnetic} indicating the existence of local heated areas in the early Universe. Hypothetical nature of local heated areas was discussed earlier \cite{dubrovich2012cosmological, kumar2019cmb, kogut2019cmb}. Such areas can appear due to large primordial density fluctuations and can be related to the clusters of Primordial Black Holes (PBHs) \cite{alihaimoud2019electromagnetic, Belotsky_2019, Belotsky:2020jyl}.

We assume that the baryonic matter has been captured by the gravitational forces of these regions at the early Universe. They would remain hot for a long time. At the same time, many processes can heat or cool the matter inside them  during their formation after it. Short list of them is the neutrino cooling \cite{Belotsky:2020jyl}, nuclear reactions, radiation of the hot plasma and stars formed inside the region \cite{Dolgov_2017}, gravitational dynamics of the system, shock waves, diffusion of matter, variation of the vacuum state while the region is born \cite{Belotsky:2017txw}, energy transfer from collapsing walls \cite{BEREZIN198391,khlopov1998formation,rubin2000primordial,Deng_2018}, accretion, the Hawking evaporation. The last mechanisms are relevant in the case of PBHs origin of the regions \cite{Belotsky_2019, dolgov1993baryon,Dolgov_2018, garcia2018primordial}. In this proceedings, we continue our consideration of neutrino cooling of such regions. It could be the most important reason for the temperature evolution within initial temperature range -- keV$<T<10$ MeV.

In this research, we follow the initial conditions taken from \cite{Belotsky_2019, Belotsky:2020jyl}, where the mass of trapped matter is in wide range $10^4$--$10^8 M_{\odot}$. The main initial parameters are as follows: the size of the region is about $R\sim 1$ pc, its mass $10^4 \, M_{\odot}$, initial temperature is in the interval $T_0\sim 1\text{ keV}\div 10$ MeV. This temperature of such regions could be reached in several ways. 
The region can start to be formed at higher temperature and finish to do it having cooled down to $T_0$.
Also, the region could be heated up during formation, e.g., in the framework of model with collapsing domain walls \cite{Kurakin:2020qqi}.

%
Without specific assumptions, we show that effect of neutrino cooling is wide spread phenomena valid in wide range of parameters.
The range of initial parameters is under consideration.

Neutrino cooling effect can be suppressed at high temperatures and large sizes when the area becomes opaque to the neutrinos.

Neutrinos are produced due to reactions of $p\leftrightarrow n$ transition and $e^+e^-$ annihilation. The characteristic time for photons to escape \textbf{the area} is bigger than the modern Universe age, this indicates that the size of cluster is big enough not to lose photons.

In the given proceedings we study the impact of the size of the \textbf{region} on the neutrino cooling effect. 

Mechanism of neutrino cooling rates for the main reactions of the neutrino production is considered in Section 2. The impact of the diffusive character of particle propagation inside the cluster is briefly discussed in Section 3.

\newpage

\section{Cooling Rates}

Let us consider the reactions of the neutrino production:
\begin{equation}
e^- + p \rightarrow n + \nu_e,
\label{ep}
\end{equation}
\begin{equation}
e^+ + n \rightarrow p + \bar\nu_e,
\label{en}
\end{equation}
\begin{equation}
e^+ + e^- \rightarrow \nu_{e,\mu,\tau}+\bar{\nu}_{e,\mu,\tau},
\label{ee}
\end{equation}
\begin{equation}
n \rightarrow p+e^- + \bar\nu_e. \label{n}
\end{equation}

The produced neutrinos leave the heated area if it is not very big. The energy inside the volume is decreased that leads to the temperature decreasing.
The rates per unit volume, $\gamma_i\equiv \Gamma_i/V$, for reactions listed above are respectively
\begin{eqnarray}
\label{gamma}
\gamma_{ep}=n_{e^-}n_p\sigma_{ep}v,\;\;\;\;\;\gamma_{en}=n_{e^+}n_n\sigma_{en}v,\\
\gamma_{ee}=n_{e^-}n_{e^+}\sigma_{ee}v,\;\;\;\;\;\;\;\;\;\;\;\;\;\;\;\;\;\;\gamma_n=\frac{n_n}{\tau_n}.
\end{eqnarray}
Here $n_i$ is the concentration of the respective species, $\sigma_{ij}$ is the cross section (see e.g. \cite{Belotsky:2020jyl}) of interacting particles $i\, {\rm and }\,j$ and $\tau_n\approx 1000$ s is the neutron lifetime. We consider the relativistic plasma so that the relative velocity $v\simeq 1$.

The backward reactions for Eqs.\eqref{ep}--\eqref{n} are suppressed if neutrinos freely escape the cluster. We consider all densities to be independent of the space coordinate inside the region. The number densities are roughly described by the following formulas, see \cite{Belotsky:2020jyl}, 
\begin{eqnarray}
n_{e^-}=n_{e^+}
+\Delta n_e,\;\;\;\;\;n_{e^+}= n_e^{eq}(T) \exp\left(-\frac{m_e}{T}\right),\label{ne}\\
n_B\equiv n_p+n_n=g_B\, \eta n_{\gamma}(T_0),\;\;\;\;\; \Delta n_e\equiv n_{e^-}-n_{e^+}=n_p.
\label{nd}
\end{eqnarray}
which are slightly corrected for 
better adjustment to the non-relativistic limit.
Here  
$\eta=n_B/n_{\gamma}\approx 0.6\cdot 10^{-9}$ {is} the baryon to photon ratio in the modern Universe, $g_B\sim 1$ is the correction factor of that relation due to entropy re-distribution, $n_{\gamma}(T)=\frac{2\zeta(3)}{\pi^2}T^3$ and $n_e^{eq}(T)=\frac{3\zeta(3)}{2\pi^2}T^3$ are the equilibrium photon and relativistic electron number densities respectively. 

Note that $n_{\gamma}(T_0)$ defines baryon density which is supposed to be unchanged starting from initial temperature $T_0$ contrary to that of $e^{\pm}$ and $\gamma$. 
Number of $e^{\pm}$ (along with $\gamma$) changes due to $e-\nu$-conversion processes (reactions Eqs.\eqref{ep} -- \eqref{n}). The temperature of the system decreases due to neutrino escape. 
Number densities of the electrons and photons fall down with temperature as $\sim T^3$. 

\section{Escaping Time}

The escape time of neutrinos from the region of the size 
$R$ with temperature $T$ 
can be calculated as:
\begin{equation}\label{t}
t_{esc}\sim \frac{R^2}{D} \sim
R^2\cdot n_e \sigma_{\nu}
\end{equation}
 in diffusion approximation.
Here the diffusion coefficient is  $D=\frac{\lambda_{\nu}\cdot v}{3}$ \cite{lifshitz2006physical}, velocity $v=1$, 
the neutrino mean free path is $\lambda_{\nu}= 1/n_e \sigma_{\nu}$ and $e-\nu$ interaction cross section was roughly taken as $\sigma_{\nu} \sim G_F^2 \cdot T^2$. The electron number density $n_e\sim n_{e^-}+n_{e^+}\sim n_{e^+}$ is given by Eq.\eqref{ne} which is $\sim T^3$ at $T>m_e$. 

One can conclude from inequality
\begin{equation}
t_{esc}\sim R^2 G_F^2 T^5<t_U
\end{equation}
that the neutrino cooling effect is significant up to the present time $t_U$ for the region of the size
\begin{equation}
R< 35 \cdot (T/{\rm MeV})^{-5/2}\,{\rm pc}.
\end{equation}
Here conditions $T\gtrsim m_e$ and $n_e\sim T^3$ are assumed.

\begin{figure}[H]
    \subfigure{
    \includegraphics[width=0.49\textwidth]{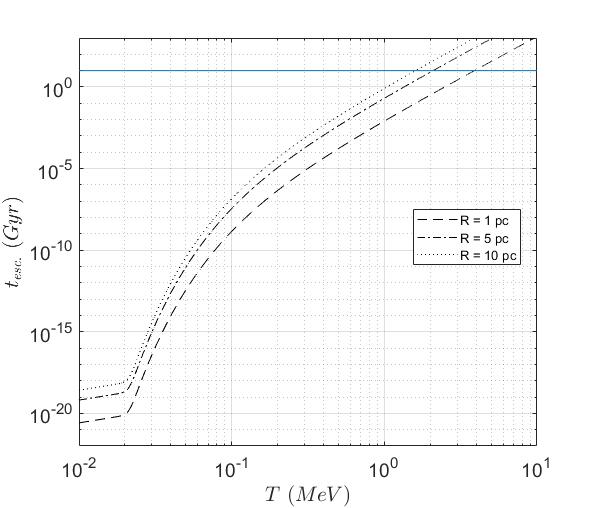}}
    \subfigure{
    \includegraphics[width=0.49\textwidth]{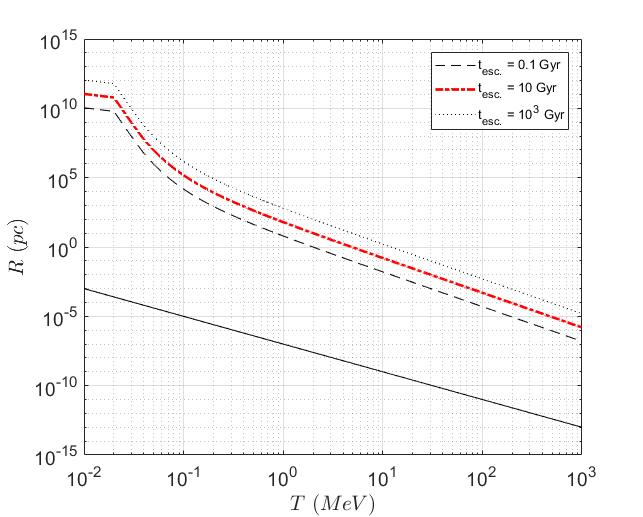}}
    \caption{Left: The relation between escaping time of neutrino and temperature of the area and the blue line is the modern age of the Universe. (Behaviour of the curves at $t_{esc}\sim 10^{-20}$ Gyr reflects the fact that $n_e$ becomes $\sim \Delta n_e$, i.e. constant.) Right: The relation between size and temperature of the area. Neutrino cooling effect plays a prominent role below the thick red dot-dashed curve. The black solid line corresponds to the dependence of the Universe horizon from its temperature ($R=10^{-7} (\text{MeV}/T)^2 \text{ pc}$).} 
    \label{fig}
\end{figure}

Neutrino cooling effect due to reactions of weak $p\leftrightarrow n$ transitions and $e^{\pm}$ annihilation are represented in Figure \ref{fig} where the escaping time of neutrinos in dependence on temperature is shown. As seen, at the temperature $T\lesssim 3$ MeV the escaping time for the most of considered cluster sizes is less than the modern Universe age, thus neutrino cooling works.
Note, that at $T\ll m_e$ the curves start to fall until the number density of electrons becomes $n_e \sim \Delta n_e$.

Dependence $R(T)$ is shown in Figure \ref{fig}, right panel, which follows from Figure \ref{fig}, left panel. The region above the red line relates to the case when neutrino cooling is suppressed (neutrinos do not run away from the region during the Universe age). Black line shows the horizon size of the Universe in dependence on the matter temperature. One can see, horizon size is much smaller than the maximal size of region at the same temperature when neutrino cooling effect is, shown by the red line. 
Therefore, the neutrino cooling effect holds under usual conditions, and can be suppressed in extreme cases.

The region can start its formation at very high temperature so that it could be cooled to the considered temperature during its detachment from Hubble flow and virialization.
Also, the region could be heated up additionally during its formation, e.g. due to wall collapsing \cite{Kurakin:2020qqi}. 
During detachment and virialization, the region could expand to some extent and hence, cool down. 

\section{Conclusions}

In earlier work \cite{Belotsky:2020jyl}, we have shown that due to neutrino emission (at a fixed size) the primordial hot areas are cooling down to the temperature value $\sim$ 0.01 $\div$ $0.1$ MeV. Here we just investigated the neutrino cooling mechanism of the heated region and dependence on its size. Considering the result of equation \eqref{t} for escaping time, we can find the size-changing more effectively with temperature. At the temperature $T > 3$ MeV, the diffusive character of particle propagation makes the time of escaping or time of cooling more than the modern Universe age. This result is obtained at the definite initial region parameters (size and temperature, relevant for PBH cluster model \cite{Belotsky_2019}) that could be slightly varied. It illustrates general property for such possible primordial inhomogeneities.

It is seen that neutrino cooling effect should take place for a wide reasonable size/temperature range of parameter. Extreme heating up of the area while it has being formed could change situation.

There are a variety of mechanisms that can be responsible for the area heating. Additional heating during their creation is also possible.
As was mentioned in the Introduction, the area could be heated by the collapsing walls - the scalar field kinks. The fermion reflection on kinks was studied in  \cite{Kurakin:2020qqi}. It was shown that the reflection weakly depends on the fermion mass. Therefore the kinks could prevent the neutrinos from escaping. More detailed analysis is necessary to clarify this effect.

\section*{Acknowledgements}
The work of K.B. was supported by the Ministry of Science and Higher Education of the Russian Federation
by project No 0723-2020-0040 ``Fundamental problems of cosmic rays and dark matter'' and of S.R. by the project "Fundamental properties of elementary particles and cosmology" No 0723-2020-0041. The work of S.R. is performed according to the Russian Government Program of Competitive Growth of Kazan Federal University and RFBR grant 19-02-00930.

\end{document}